\documentclass[pdflatex, sn-mathphys]{sn-jnl}

\usepackage{braket}
\usepackage{amsmath}
\usepackage{amsmath, bm}
\usepackage{physics}
\usepackage{upgreek}
\usepackage[mathlines]{lineno}

\renewcommand{\bra}[1]{\langle#1\vert} 
\renewcommand{\ket}[1]{\vert#1\rangle} 
\renewcommand{\braket}[1]{\langle{#1}\rangle} 

\usepackage{etoolbox} 
\newcommand*\linenomathpatch[1]{%
  \cspreto{#1}{\linenomath}%
  \cspreto{#1*}{\linenomath}%
  \csappto{end#1}{\endlinenomath}%
  \csappto{end#1*}{\endlinenomath}%
}
\linenomathpatch{equation}
\linenomathpatch{gather}
\linenomathpatch{multline}
\linenomathpatch{align}
\linenomathpatch{alignat}
\linenomathpatch{flalign}

\jyear{2023}
\theoremstyle{thmstyleone}

\theoremstyle{thmstyletwo}%
\theoremstyle{thmstylethree}%
\setcitestyle{sort&compress,open={},close={},numbers,super} 

\begin{document}

\title[Minutes-scale Schr{\"o}dinger-cat state of spin-5/2 atoms]{Minutes-scale Schr{\"o}dinger-cat state of spin-5/2 atoms}

\author[1]{\fnm{Y. A.} \sur{Yang}}\email{yyustcer@mail.ustc.edu.cn}

\author[1]{\fnm{W.-T.} \sur{Luo}}\email{lspecwave@mail.ustc.edu.cn}

\author[1]{\fnm{J.-L.} \sur{Zhang}}\email{zjl\_2001@mail.ustc.edu.cn}

\author[1]{\fnm{S.-Z.} \sur{Wang}}\email{szwang@mail.ustc.edu.cn}

\author[1,2,3]{\fnm{Chang-Ling} \sur{Zou}}\email{clzou321@ustc.edu.cn}

\author*[1]{\fnm{T.} \sur{Xia}}\email{txia1@ustc.edu.cn}

\author*[1,2]{\fnm{Z.-T.} \sur{Lu}}\email{ztlu@ustc.edu.cn}

\affil[1]{\orgdiv{CAS Center for Excellence in Quantum Information and Quantum Physics, School of Physical Sciences}, \orgname{University of Science and Technology of China}, \orgaddress{ \city{Hefei}, \postcode{230026}, \state{Anhui}, \country{China}}}
\affil[2]{\orgdiv{Hefei National Laboratory}, \orgname{University of Science and Technology of China}, \orgaddress{\city{Hefei}, \postcode{230088}, \state{Anhui}, \country{China}}}

\affil[3]{\orgdiv{CAS Key Laboratory of Quantum Information}, \orgname{University of Science and Technology of China}, \orgaddress{\city{Hefei}, \postcode{230026}, \state{Anhui}, \country{China}}}

\abstract{Quantum metrology with nonclassical states offers a promising route to improved precision in physical measurements\cite{Pezze2018}. The quantum effects of Schr{\"o}dinger-cat superpositions \cite{Raimond2006,Monroe1996a,Facon2016,Chalopin2018,Dietsche2019,Bild2023} or entanglements\cite{Pezze2018,Gross2010,Strobel2014,Hosten2016,Bao2020, Pedrozo-Penafiel2020, Erhard2020,Liu2022,Greve2022} allow measurement uncertainties to reach below the standard quantum limit\cite{Caves1980}. However, the challenge in keeping a long coherence time for such nonclassical states often prevents full exploitation of the quantum advantage in metrology. Here we demonstrate a long-lived Schr{\"o}dinger-cat state of optically trapped $^{173}$Yb (\textit{I}\ =\ 5/2) atoms. The cat state, a superposition of two oppositely-directed and furthest-apart spin states, is generated by a non-linear spin rotation. Protected in a decoherence-free subspace against inhomogeneous light shifts of an optical lattice, the cat state achieves a coherence time of $1.4(1)\times 10^3$ s. A magnetic field is measured with Ramsey interferometry, demonstrating a scheme of Heisenberg-limited metrology for atomic magnetometry, quantum information processing, and searching for new physics beyond the Standard Model\cite{Safronova2018, Chupp2019}.
}


\maketitle

\section{Introduction}\label{sec1}

Quantum metrology employs nonclassical states, including Schr{\"o}dinger cat states\cite{Raimond2006,Monroe1996a,Facon2016,Chalopin2018,Dietsche2019,Bild2023} and entangled states\cite{Pezze2018,Gross2010,Strobel2014,Hosten2016,Bao2020, Pedrozo-Penafiel2020, Erhard2020,Liu2022,Greve2022}, to improve the measurement precision beyond the standard quantum limit (SQL)\cite{Caves1980}. For example, a cat state of Rydberg atoms acted as a sensitive probe of electric or magnetic fields\cite{Facon2016,Dietsche2019}; entangled states of many particles outperformed classical states of uncorrelated particles in an interferometric phase measurement on a Bose-Einstein condensate\cite{Gross2010,Pezze2018}. A cat state of single particles is compatible and can be combined with multi-particle entanglement to further improve the precision in quantum metrology. However, utilizing nonclassical states in measurements faces two key challenges: first, preparing them commonly requires nonlinear or nonlocal operations\cite{Raimond2001, Pezze2018}; second, preserving the coherent nonclassical states for long is especially difficult due to their fragility in the environmental\cite{Zurek2003, Frowis2018}. 

In an interferometric-phase measurement on either a particle of spin $F$ or an ensemble of $2F$ particles of spin 1/2, the SQL is $\Delta\phi_{\mathrm{SQL}}=1/\sqrt{2F}$, while the lower Heisenberg limit (HL) $\Delta\phi_{\mathrm{HL}}=1/(2F)$ can be approached with nonclassical quantum states including Schr{\"o}dinger-cat states\cite{Facon2016}, NOON states\cite{Walther2004,  Mitchell2004} and squeezed states\cite{Hald1999, Orzel2001}. In practice, these nonclassical states often suffer from short coherence times\cite{Zurek1991, Zurek2003, Frowis2018}. However, decoherence\cite{Myatt2000, Deleglise2008} is not an intrinsic property of a quantum system. Quantum decoherence effects and their reduction have been extensively studied\cite{Lidar1998, Zurek2003, Haroche2013}. A quantum system embedded in a decoherence-free subspace of the overall Hilbert space undergoes collective decoherence, and thus is insensitive to certain types of noise\cite{Lidar1998}. For example, a two-ion Bell state insensitive to first-order Zeeman shifts has been demonstrated\cite{Kielpinski2001, Roos2006}. Moreover, compared with a spin-$1/2$ particle, a spin-$F$ particle benefits from an amplified intrinsic sensitivity, and the redundancy of the high-dimensional Hilbert space in a spin-$F$ particle allows for advanced error suppression techniques such as the decoherence-free subspace\cite{Lidar1998, Kielpinski2001, Roos2006} and quantum error correction.

\begin{figure}[!htb]
\centering
\includegraphics[width=1.0\textwidth]{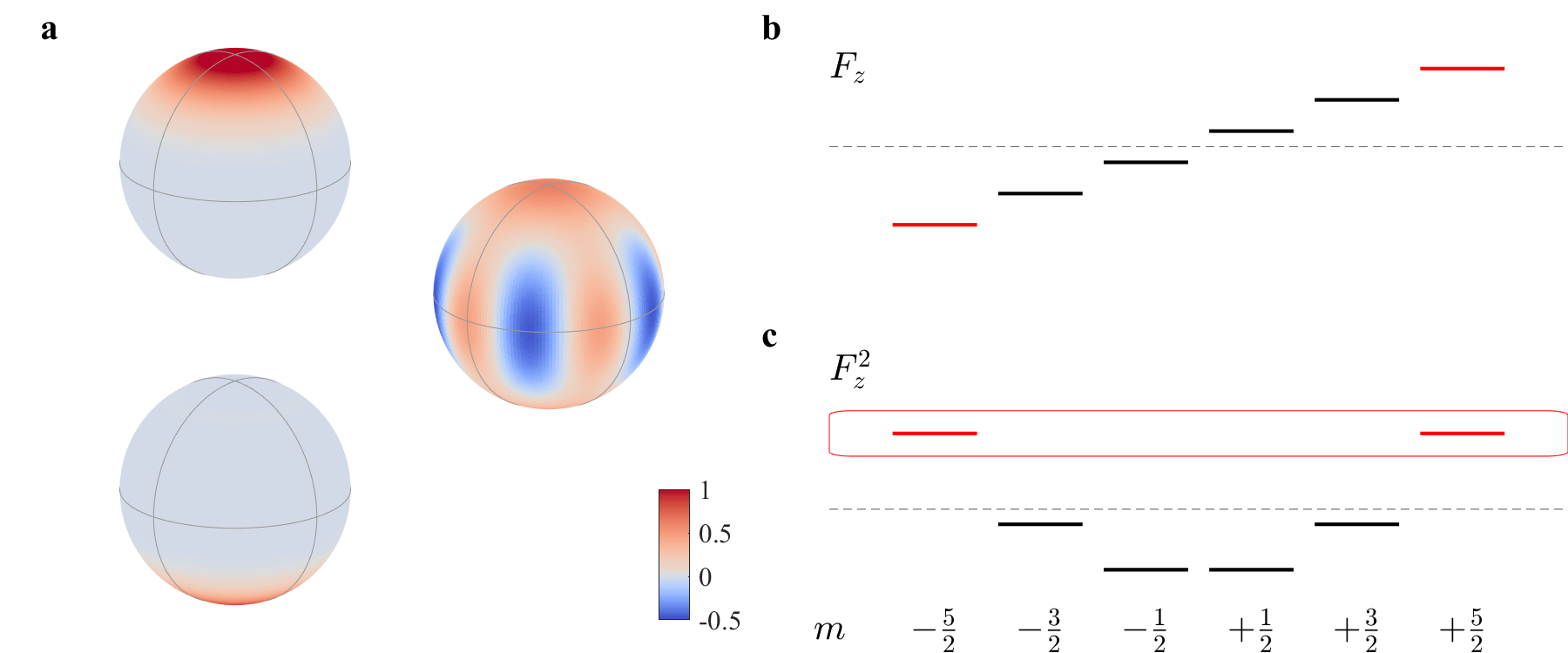}
\caption{\textbf{Principle of the method.} 
\textbf{a}, Wigner functions of the $m=+5/2$ state (upper left), -5/2 state (lower left), and their equal superposition (right) forming the Schr{\"o}dinger-cat state of a spin-5/2 system. The negative-valued Wigner function indicates nonclassicality, and displays a $2\pi/(2F)$ rotation symmetry around the quantization $z$-axis.
\textbf{b}, Zeeman shifts ($\delta E\propto m$) of a spin-5/2 system induced by $F_z$. The cat state is the superposition of two stretched states indicated by red lines.
\textbf{c}, Quadratic energy shifts ($\delta E\propto m^2$) induced by a tensor interaction $F_z^2$. The interferometric phase of the cat state, encoded in the subspace $\mathcal{H}_{\pm 5/2}$ as indicated by the red box, is insensitive to variations of the quadratic shifts. This scheme applies to spins of arbitrary sizes.
}\label{fig1}
\end{figure}

Here we demonstrate a long-lived Schr{\"o}dinger-cat state using spin-5/2 nuclei of $\mathrm{^{173}Yb}$ atoms trapped in an optical lattice. The cat state, a quantum superposition of the $m=+5/2$ and $m=-5/2$ nuclear-spin projection states (Fig.~\ref{fig1}\textbf{a}), is prepared by a pulse of control laser to induce nonlinear light shifts on the ground states. Protected within a decoherence-free subspace, the cat state is insensitive to inhomogeneous tensor light shifts induced by the lattice trap laser (Fig.~\ref{fig1}\textbf{c}), and achieves a long coherence time of $1.4(1)\times 10^3$ s. Ramsey interferometry is performed to demonstrate a measurement sensitivity approaching the Heisenberg limit. The cat state exhibits a maximal sensitivity to the bias magnetic field (Fig.~\ref{fig1}\textbf{b}) and potential vector shifts induced by new physics beyond the Standard Model, such as in searches for a permanent electric dipole moment, tests of the Lorentz invariance, and searches for exotic spin-dependent interactions\cite{Safronova2018, Chupp2019}. In addition, a nuclear-spin state with a long coherence time is also an attractive qubit candidate\cite{Jenkins2021a, Ma2022,Huie2023}. The cat state of single particles is compatible with schemes of multi-particle entanglement, suggesting room for further improvement in quantum metrology by combining both approaches. 

\section{Heisenberg-limited sensitivity}\label{sec2}

In this work, the spin-$F$ system is the $\mathrm{^{173}Yb}$ atom in the ground state $\mathrm{6s^2\ ^1S_0}$ ($F=5/2$). The energy splitting generated by a magnetic field $B_z$ (Fig.~\ref{fig1}\textbf{b}) can be measured with Ramsey interferometry on the coherent spin state (CSS) $\ket{\theta,\varphi}$ with $\theta=\pi/2$, where $\ket{\theta,\varphi}=\mathrm{e}^{-\mathrm{i}\varphi F_z}\mathrm{e}^{-\mathrm{i}\theta F_y}\ket{F, F}$, and $F_{x,y,z}$ are the angular momentum operators\cite{Landauer1972}. The best sensitivity achievable is set by the quantum Cram{\'e}r-Rao bound $\Delta \phi\geq \frac{1}{\sqrt{\mathcal{F}}}$ on the phase shift, where $\mathcal{F}=\mathrm{4 Var(\mathit{F_z})}$ is the quantum Fisher information of the state\cite{Pezze2018,Kwon2019}. For the CSS $\ket{\pi/2,\varphi}$, $\mathcal{F}_{\mathrm{CSS}}=2F$, and the sensitivity limit $1/\sqrt{2F}$ is known as the standard quantum limit (SQL)\cite{Caves1980,Facon2016}. On the other hand, the Schr{\"o}dinger-cat state $\ket{\psi_{\pm F}}=1/\sqrt{2}(\ket{F, F}+\mathrm{e}^{\mathrm{i}\phi_{F}}\ket{F, -F})$ provides the maximum possible $\mathcal{F}_{\mathrm{cat}}=4F^2$, and can achieve the Heisenberg-limited sensitivity of $1/(2F)$. As shown in Fig.~\ref{fig1}\textbf{a}, the cat state 
displays a $2\pi/{(2F)}$ rotation symmetry around $F_{z}$ in its negative-valued Wigner function\cite{Dowling1994}, corresponding to a $2F$-fold enhancement of the repetition frequency over CSS (Fig.~\ref{fig2}\textbf{e}). As a result, the magnetic-field sensitivity of a single measurement set by the Cram{\'e}r-Rao bound is 
$\sigma_{B}^{1}=\frac{1}{\sqrt{\mathcal{F}}}\frac{1}{\gamma \tau}$, 
where $\gamma$ and $\tau$ are the gyromagnetic ratio and interrogation time, respectively. 
 
 \section{Generation of the cat state}\label{sec3}
\begin{figure}[!htb]
\centering
\includegraphics[width=1.0\textwidth]{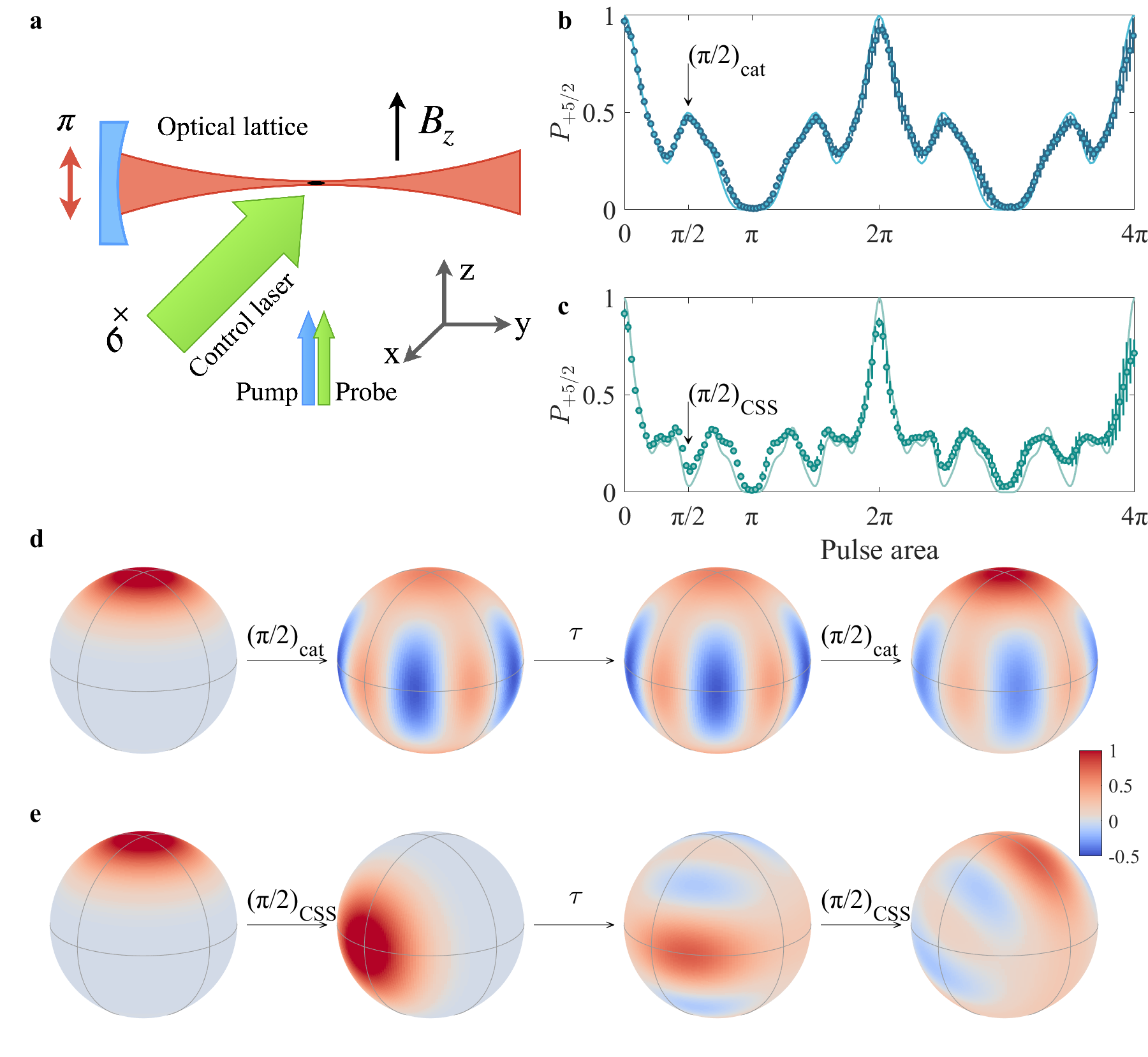}
\caption{\textbf{Setup and spin dynamics.} 
\textbf{a}, Schematic of the setup. $\mathrm{^{173}Yb}$ atoms are trapped in an optical lattice. Both the bias magnetic field and the linear polarization of the lattice are along the $\hat{z}$ direction. The pump, probe, and control laser beams, all $\sigma^{+}$-polarized, are for initialization, state-selective measurement, and coherent control of the spin, respectively.
\textbf{b}, Normalized population $P_{+5/2}$ in the $m=+5/2$ state as a function of the control-pulse area. The data points match the simulated results (solid line). A $(\pi/2)_{\mathrm{cat}}$ pulse (marked by the arrow) is used to coherently drive the initial state $\ket{F, F}$ to the cat state.  
\textbf{c}, A $(\pi/2)_{\mathrm{CSS}}$ pulse drives $\ket{F, F}$ to the coherent spin state $\ket{\pi/2,\varphi}$. 
\textbf{d}, Ramsey sequence with the cat state. The initial state $\ket{F, F}$ is brought to the cat-state superposition by a $(\pi/2)_{\mathrm{cat}}$ optical pulse. A second $(\pi/2)_{\mathrm{cat}}$ pulse recombines the superposition, and the normalized populations of $\ket{F, \pm F}$ are measured. 
\textbf{e}, Ramsey sequence with the coherent spin state (CSS). The initial state $\ket{F, F}$ is brought to the CSS $\ket{\pi/2,\varphi}$ by a $(\pi/2)_{\mathrm{CSS}}$ optical pulse. The CSS would evolve to a mixture of superposition states during interrogation due to the nonuniform tensor light shifts of the lattice. Error bars in \textbf{b} and \textbf{c} represent 1 standard deviation, and some are smaller than the symbol size.
}\label{fig2}
\end{figure}

$^{173}$Yb atoms are trapped in an optical lattice linearly polarized in the $z$-direction (Fig.~\ref{fig2}\textbf{a}). The 16-W lattice beam has a waist of 20 \textmu m, resulting in a trap depth of 2.4 mK. It operates at 1036 nm, the magic wavelength for the $\mathrm{6s^2\ ^1S_0}\rightarrow\mathrm{6s6p\ ^3P_1}$ transition\cite{Zheng2020}. The atoms are first loaded from a magneto-optical trap in a neighboring chamber, and then transported to the measurement chamber by a moving optical dipole trap directed along the $\hat{x}$-axis. On the order of $10^{4}$ atoms are handed over to the optical lattice. The beam waist of the moving trap, at 70 \textmu m, determines the length over which the atoms occupy along the optical axis of the lattice. The vacuum trap lifetime of atoms in the lattice is 71(1) s. The apparatus is protected by four layers of magnetic shielding, within which a cos($\theta$) coil generates a stable and uniform magnetic field of 1.24 µT along the $\hat{z}$ direction.

Protocols for producing nonclassical states generally require an effective nonlinear interaction\cite{Yurke1986, Sanders1989}, which, in this work, is the $m^2$-dependent tensor ac Stark shifts\cite{Smith2004, Chalopin2018}. The spin control and measurement take the following steps. First, a pulse of $\sigma^{+}$-polarized pump laser beam along the $\hat{z}$ direction is applied to initialize the atoms into the $\ket{F, F}$ stretched state. The pump laser is on resonance with the $\mathrm{^1S_0}(F=5/2)\rightarrow\mathrm{^1P_1}(F'=5/2)$ non-cycling transition\cite{Zheng2022}. Second, a pulse of off-resonant control laser beam is applied to drive a spin rotation. The $\sigma^{+}$-polarized control laser beam propagating along $\hat{x}$ (Fig.~\ref{fig2}\textbf{a}) interacts with the atoms according to the Hamiltonian $H_{\mathrm{c}}= \mathit{\Omega}_{x}^{(1)} F_x+\mathit{\Omega}_{x,x}^{(2)} F_x^2$, where $\mathit{\Omega}_{x}^{(1)}$ and $\mathit{\Omega}_{x,x}^{(2)}$ are the generalized Rabi frequencies (see Methods). The ratio $\mathit{\Omega}_{x}^{(1)}/\mathit{\Omega}_{x,x}^{(2)}$ can be set within a wide range by tuning the frequency of the control laser between the hyperfine resonances of the $\mathrm{^1S_0}\rightarrow\mathrm{^3P_1}$ transition. Third, the normalized population remaining in $\ket{F, F}$,  $P_{+F}$, is measured with the $\sigma^{+}$-polarized probe beam (Fig.~\ref{fig2}\textbf{a}). The probe beam is resonant with the $\mathrm{^1S_0}(F=5/2)\rightarrow\mathrm{^3P_1}(F'=7/2)$ cycling transition where the differential tensor light shifts induced by the optical lattice enable state-selective measurements.

In the case of $\vert\mathit{\Omega}_{x}^{(1)}/\mathit{\Omega}_{x,x}^{(2)}\vert=1$, the initial state $\ket{F, F}$ is coherently driven to the cat state $1/\sqrt{2}(\ket{F, F}-i\ket{F, -F})$ after applying a control pulse $(\pi/2)_{\mathrm{cat}}$ with a duration $t=\pi/(2\vert \mathit{\Omega}_{x}^{(1)}\vert)$ (Fig.~\ref{fig2}\textbf{b}). The light intensity of the control laser beam is approximately 80 mW/cm$^{2}$, which gives a measured Rabi frequency of $\mathit{\Omega}_{x}^{(1)}=-\mathit{\Omega}_{x,x}^{(2)}=2\pi\times0.5187(6)\ \mathrm{kHz}$. This $(\pi/2)_{\mathrm{cat}}$ pulse is analogous to a $\pi/2$ pulse in a Ramsey method on a spin-1/2 system. A separated measurement on a coherent spin state (CSS) is also performed. Here the CSS $\ket{\pi/2,\varphi}$ is prepared by a $(\pi/2)_{\mathrm{CSS}}$ control laser pulse where $\vert\mathit{\Omega}_{x}^{(1)}/\mathit{\Omega}_{x,x}^{(2)}\vert=1/2$ (Fig.~\ref{fig2}\textbf{c}) (see Methods). The periodic but non-sinusoidal waveforms in Fig.~\ref{fig2}\textbf{b} and Fig.~\ref{fig2}\textbf{c} are the results of rotation operations on a high-spin.

 \section{Decoherence-free cat state}\label{sec4}
 
Inhomogenous light shift, usually the dominant source of decoherence in optical traps, can greatly limit the coherence time and cause systematic errors. The application of `magic wavelength'\cite{Ye2008} or `magic angle'\cite{Neyenhuis2012} can effectively eliminate the effects of such light shift in some atomic or molecular systems. Here we show that the cat state is protected in a decoherence-free subspace. The Hamiltonian for spin-light interaction in the optical lattice is given by
\begin{equation}
H_{\mathrm{t}}(I)=\widetilde{\mathit{\Omega}}_{z,z}^{(2)}F_z^2,
\end{equation}
where $\widetilde{\mathit{\Omega}}_{z,z}^{(2)}$ is the generalized tensor Rabi frequencies determined by the intensity distribution $p(I)$ of the trap light field. The off-diagonal terms in the total Hamiltonian $H(I)=H_0+H_{\mathrm{t}}(I)$, where $H_0=\gamma B F_z =\mathit{\Omega}_0 F_z$, are all equal to zero, thus blocking leakage to intermediate states. The density matrix after interrogation time $\tau$ can be calculated as an average over all possible $I$. In particular, $H_{\mathrm{t}}$ commutes with both the cat-state density matrix $\rho_{\mathrm{cat}}$ and $H_0$, then $\rho_{\mathrm{cat}}(\tau)=\int_{I} \mathrm{e}^{-\mathrm{i} H(I)\tau/\hbar}\rho_{\mathrm{cat}}(0)\mathrm{e}^{\mathrm{i} H(I)\tau/\hbar}p(I)\mathrm{d}I=\int_{I} \mathrm{e}^{-\mathrm{i} H_0\tau/\hbar}\rho_{\mathrm{cat}}(0)\mathrm{e}^{\mathrm{i} H_0\tau/\hbar}p(I)\mathrm{d}I$. Thus the decoherence of the cat state brought by the light field is avoided and the cat state is protected in a decoherence-free subspace $\mathcal{H}_{\pm F}$ (\{$\ket{F, F}$, $\ket{F, -F}$\}). The Heisenberg-limited sensitivity $1/(2F)$ can be maintained for a long time since there is no reduction of quantum Fisher information.

For comparison, the CSS $\ket{\pi/2,\varphi}$ would evolve to a mixture of superposition states under the same condition. Since $\ket{\pi/2,\varphi}$ spans over all possible decoherence-free subspaces $\mathcal{H}_{\pm m}$ (\{$\ket{F, m}$, $\ket{F, -m}$\}), $H_{\mathrm{t}}$ results in dephasing between these subspaces and $\ket{\pi/2,\varphi}$ evolves into a mixture $\rho_{\mathrm{mix}}=\sum_{\pm m }w_{\pm m} \ket{\psi_{\pm m}} \bra{\psi_{\pm m}}$, where $\ket{\psi_{\pm m}}=1/\sqrt{2}(\ket{F, m}+\mathrm{e}^{i\phi_{m}}\ket{F, -m})$ are the superposition states encoded in $\mathcal{H}_{\pm m}$, and $w_{\pm m}$ are the weighting factors. The timescale of the CSS-to-mixture evolution is reciprocal to the spreading of $\widetilde{\mathit{\Omega}}_{z^2}$. This mixture is also nonclassical indicated by the negative-valued Wigner function in Fig.~\ref{fig2}\textbf{e}. However, the corresponding quantum Fisher information of the mixture state is $\mathcal{F}_{\mathrm{mix}}=2F$, no greater than that of the CSS (see Methods). 

\section{Ramsey interferometry}\label{sec5}

\begin{figure}[!htb]
\centering
\includegraphics[width=1.0\textwidth]{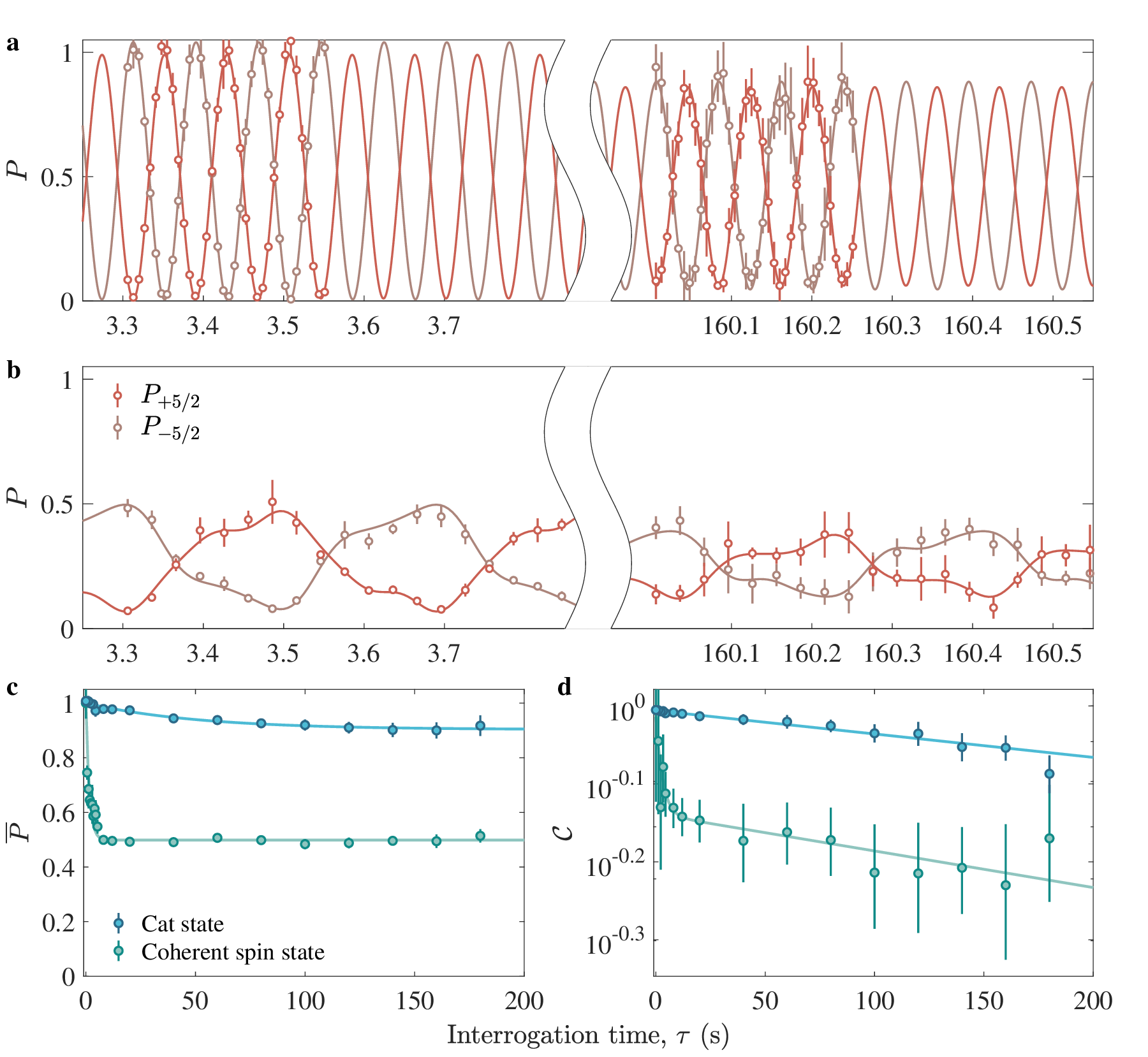}
\caption{\textbf{Ramsey interference.} 
\textbf{a}, Normalized populations of $\ket{F, \pm F}$ as a function of interrogation time $\tau$ in Ramsey interferometry using a cat state. The higher repetition frequency is a signature of quantum enhancement. 
\textbf{b}, Normalized populations of $\ket{F, \pm F}$ as a function of interrogation time $\tau$ in a CSS-based Ramsey interferometry.
\textbf{c}, Populations remaining in $\ket{5/2,\pm 5/2}$ states as a function of $\tau$. The blue points show the results of the Ramsey sequence with the cat state; the green points show those with the CSS, which collapses quickly into a mixed state. Solid lines are exponential fits.
\textbf{d}, Fringe contrast as a function of $\tau$. The coherence times extracted from exponential fits are $T_{2, \mathrm{cat}}^{*}=1.4(1) \times 10^3$ s and $T_{2, \mathrm{CSS}}^{*}=0.9(2)\times 10^3$ s. Error bars represent 1 standard deviation, sometimes smaller than the symbol size.
}\label{fig3}
\end{figure}

We perform a Ramsey interferometric measurement to characterize the sensitivity of the cat state to a 1.24-\textmu T static magnetic field\cite{Ramsey1950}. Two sequential $(\pi/2)_{\mathrm{cat}}$ pulses separated by an interrogation time $\tau$ are applied to the initial state $\ket{F, F}$. As shown in Fig.~\ref{fig2}\textbf{d}, the first $(\pi/2)_{\mathrm{cat}}$ pulse generates the cat state for magnetic sensing; the phase between $\ket{F, F}$ and $\ket{F, -F}$ are accumulated during the interrogation time $\tau$; the second $(\pi/2)_{\mathrm{cat}}$ pulse recombines the superposition and converts the phase accrued during $\tau$ to a population imbalance between $\ket{F, F}$ and $\ket{F, -F}$. We probe the normalized population in $\ket{F, F}$, $P_{+F}$. Then we apply a $(\pi)_{cat}$ pulse to swap the populations between $\ket{F, F}$ ($P_{+F}$) and $\ket{F, -F}$ ($P_{-F}$), and measure the original $P_{-F}$ prior to swapping. The measured $P_{+F}$ and $P_{-F}$ oscillate sinusoidally as a function of interrogation time $\tau$ (Fig.~\ref{fig3}\textbf{a}). The repetition frequency is $2F$ times that of the CSS (Fig.~\ref{fig3}\textbf{b}), demonstrating a clear quantum-enhanced signature.

\begin{figure}[!htb]
\centering
\includegraphics[width=1.0\textwidth]{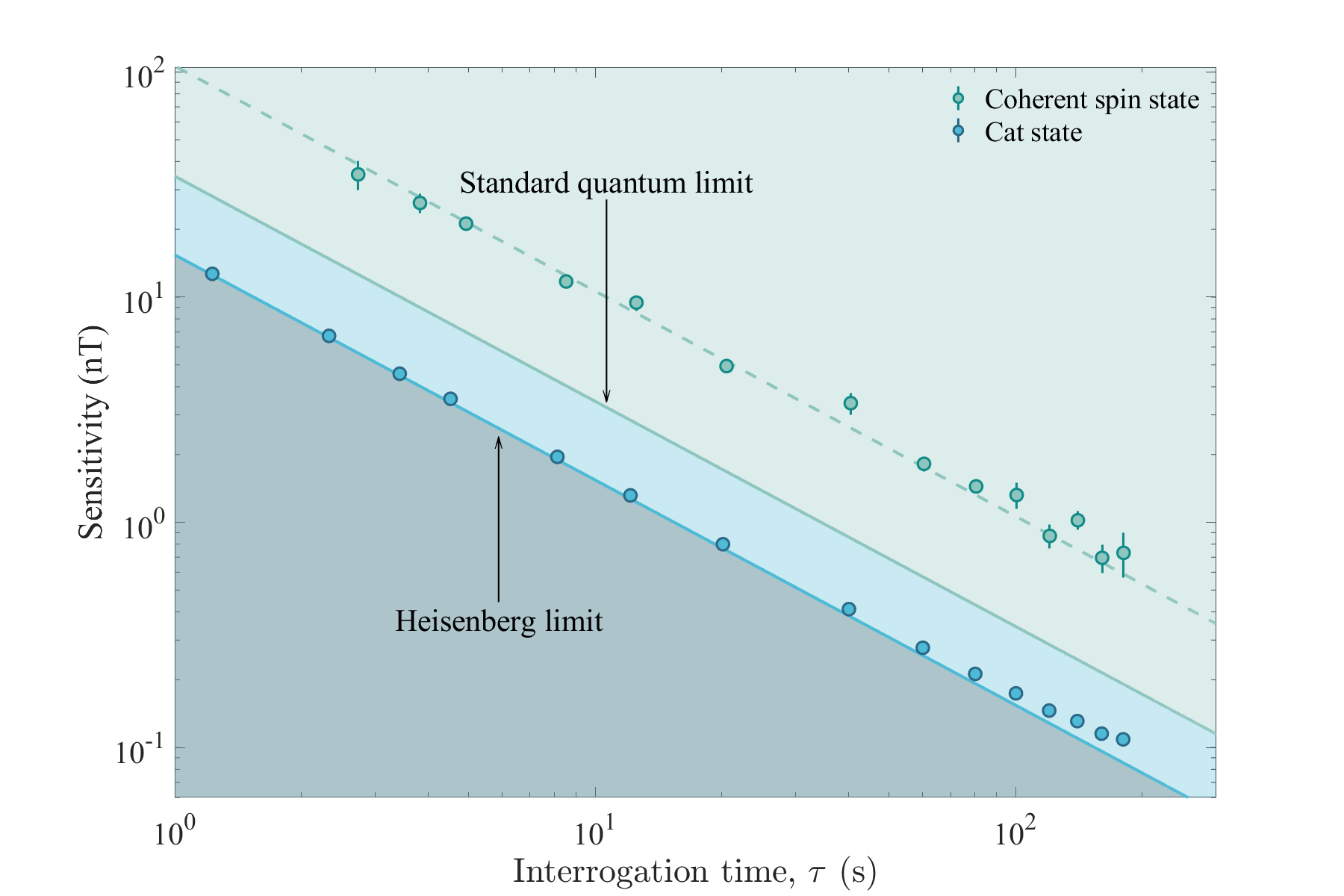}
\caption{\textbf{Sensitivity to a magnetic field.} 
The blue points show the magnetic-field sensitivity $\sigma^{1}_{B}$ of the cat state as a function of $\tau$ in a single measurement; the green points are for the CSS. The cat-state sensitivity reaches the HL over a timescale of 100 s. The sensitivity of the CSS, shown by green points, is shifted above the SQL due to the incomplete detection of all states. The dashed line represents a weighted $1/\tau$ fit to the CSS data. Error bars represent 1 standard deviation, sometimes smaller than the symbol size. 
}\label{fig4}
\end{figure}

The detrimental effects of the noisy environment, including the leakage into intermediate levels and the dephasing process, can cause loss of Fisher information and a reduced sensitivity. We extract the probability ($\overline{P}$) of states remaining in $\mathcal{H}_{\pm F}$ and the fringe contrast ($\mathcal{C}$) by fitting $P_{\pm F}$ with the following function
\begin{equation}
\label{cosinefit}
P(\tau)=\frac{\overline{P}(\tau)}{2}[1+\mathcal{C}(\tau)\cos(\omega\tau+\varphi_{c})].
\end{equation}
We find $\overline{P}=0.90(3)$ and $\mathcal{C}=0.88(3)$ after an interrogation time of 160 s (Fig.~\ref{fig3}\textbf{c} and \ref{fig3}\textbf{d}). An exponential fit to the fringe contrast $\mathcal{C}$ as a function of $\tau$ (Fig.~\ref{fig3}\textbf{d}) leads to the coherence time $T_{2, \mathrm{cat}}^{*}=1.4(1) \times 10^3$ s. We evaluate the magnetic field sensitivity of the cat state with classical Fisher information (see Methods). Taking into account the losses in both the population and interference contrast, the magnetic field sensitivity by a single measurement is 
\begin{equation}
\sigma_{B}^{1}=\frac{1}{\mathcal{C}\sqrt{\overline{P}}} \frac{1}{2F} \frac{1}{\gamma \tau},
\end{equation}
where $2F=\sqrt{\mathcal{F}_{\mathrm{cat}}}$. Fig.~\ref{fig4} shows $\sigma_{B}^{1}$ as a function of $\tau$, which saturates at the Heisenberg limit of $\sigma_{B, \mathrm{HL}}^{1}=\frac{1}{2F} \frac{1}{\gamma \tau}$ over the timescale of $100$ s. The sensitivity reaches 0.12(1) nT for $\tau=160$ s, close to the Heisenberg limit (HL) of 0.10 nT, and is improved over the SQL of 0.22 nT by a factor of 1.8. We note that the Ramsey sequence on the cat state represents nonlinear interferometry\cite{Gross2010} since the preparation of the superposition employs the nonlinear rotation operator $F_x+F_x^2$.

For comparison, we replace $(\pi/2)_{\mathrm{cat}}$ pulses with $(\pi/2)_{\mathrm{CSS}}$ pulses and perform a Ramsey experiment on the CSS (Fig.~\ref{fig2}\textbf{e}). The evolution of this high-spin system is fit with three different frequencies (Fig.~\ref{fig3}\textbf{b}). $\overline{P}$ decays quickly to the value of $0.5$ (Fig.~\ref{fig3}\textbf{c}), and the fitted exponential time constant is $T_{\mathrm{CSS}}=1.5(2)$ s. Fitting $\mathcal{C}$ with a double-exponential function, we extract the coherence time $T_{2, \mathrm{CSS}}^{*}=0.9(2)\times 10^3$ s. The quantum Fisher information $\mathcal{F}_{\mathrm{mix}}$ equals the classical limit $2F$. However, it would require measuring all states in order to approach the SQL. In our situation, only $P_{+F}$ and $P_{-F}$ are measured, resulting in a reduction of the Fisher information. The sensitivity reaches 0.70(10) nT for $\tau=160$ s, about 3.2 times larger than the SQL of 0.22 nT. 

\section{Discussion}\label{sec6}

We have demonstrated a Schr{\"o}dinger-cat state of a high-spin system with a coherence time of $1.4(1) \times 10^3$ s. With better vacuum, the lifetime of the cat state can be lengthened to match the coherence time. Spin echo can be applied on the cat state to further reduce decoherence effects. In measuring external magnetic fields, the cat state exhibits a sensitivity approaching the Heisenberg limit, and is enhanced by 15(2) dB over the coherent spin states. Our method enables efficient initialization, manipulation, and detection of spin-$F$ systems. This experimental approach contributes to the realm of practical quantum metrology by leveraging the internal degrees of freedom\cite{Facon2016,Chalopin2018,Dietsche2019}. Moreover, the high-spin system can be applied to quantum memory \cite{Jenkins2021a,Ma2022,Huie2023}, and provide the redundancy needed in quantum error corrections and potential qudit encodings for quantum computations\cite{Ringbauer2022}.

An atom with a ground level of $J = 0$ and a nuclear spin $I = 1/2$ is generally considered an ideal candidate for spin sensors. The few choices available include $\mathrm{^{3}He}$, $\mathrm{^{129}Xe}$, $\mathrm{^{171}Yb}$, and $\mathrm{^{199}Hg}$. No spin-1/2 isotope pairs exist for any of these elements. With this work, high-spin isotopes can now join the candidate pool. $\mathrm{^{171}Yb}$/$\mathrm{^{173}Yb}$ are but one example of an isotope pair suitable for the development of a dual-species cold-atom comagnetometer.
\section{Methods}\label{sec7}

\subsection{The ac-Stark operator}

The light shifts or ac Stark shifts\cite{Rosenbusch2009, LeKien2013} of an atomic state are described by
\begin{equation}
\begin{aligned}
H_{\mathrm{ac}}=-\frac{1}{4}\vert\bm{\mathcal{E}}\vert^2 &\left\{\alpha_F^{\mathrm{S}}-i \alpha_F^{\mathrm{V}} \frac{\left[\bm{u}^* \times \bm{u}\right] \cdot \bm{F}}{2 F}\right.\\
+&\left.\alpha_F^{\mathrm{T}} \frac{3\left[\left(\bm{u}^* \cdot \bm{F}\right)(\bm{u} \cdot \bm{F})+(\bm{u} \cdot \bm{F})\left(\bm{u}^* \cdot \bm{F}\right)\right]-2 \bm{F}^2}{2 F(2 F-1)}\right\}
\end{aligned}
\end{equation}
where $\alpha_F^{\mathrm{S}}$, $\alpha_F^{\mathrm{V}}$, and $\alpha_F^{\mathrm{T}}$ are the dynamical scalar, vector, and tensor polarizabilities of the state with a total angular momentum $\bm{F}\equiv(F_x, F_y, F_z)$, respectively; $\bm{\mathcal{E}}$ and $\bm{u}\equiv(u_x,u_y,u_z)$ are the complex amplitude and unit polarization vector of the light field, respectively. The Hamiltonian can also be expressed in terms of angular momentum operators $(F_x, F_y, F_z)$ as  
\begin{equation}
H_{\mathrm{ac}}=\mathit{\Omega}^{\mathrm{S}}+\sum_{i}^{x,y,z}\mathit{\Omega}^{\mathrm{V}}_{i}F_{i}+\sum_{i,j}^{x,y,z}\mathit{\Omega}^{\mathrm{T}}_{ij}F_{i}F_{j}
\end{equation}
where $\mathit{\Omega}^{\mathrm{S}}$, $\mathit{\Omega}_{i}^{\mathrm{V}}$, and $\mathit{\Omega}_{i,j}^{\mathrm{T}}$ are the generalized scalar, vector, and tensor Rabi frequencies, respectively.
\begin{equation}
	\begin{aligned}
		\mathit{\Omega}^{\mathrm{S}}&=-\frac{\alpha_F^{\mathrm{S}}}{4}\vert\bm{\mathcal{E}}\vert^2 \\
		\mathit{\Omega}_{x}^{\mathrm{V}}&=\frac{i\alpha_F^{\mathrm{V}}}{8F}\vert\bm{\mathcal{E}}\vert^2(u_z u_y^*-u_y u_z^*) \\
		\mathit{\Omega}_{y}^{\mathrm{V}}&=\frac{i\alpha_F^{\mathrm{V}}}{8F}\vert\bm{\mathcal{E}}\vert^2(u_x u_z^*-u_z u_x^*) \\
		\mathit{\Omega}_{z}^{\mathrm{V}}&=\frac{i\alpha_F^{\mathrm{V}}}{8F}\vert\bm{\mathcal{E}}\vert^2(u_y u_x^*-u_x u_y^*) \\
		\mathit{\Omega}_{i,i}^{\mathrm{T}}&=-\frac{\alpha_F^{\mathrm{T}}}{4F(2F-1)}\vert\bm{\mathcal{E}}\vert^2(3u_i u_i^*-1) \\
		\mathit{\Omega}_{i,j}^{\mathrm{T}}&=-\frac{3\alpha_F^{\mathrm{T}}}{8F(2F-1)}\vert\bm{\mathcal{E}}\vert^2(u_i u_j^*+u_i^* u_j). \\
	\end{aligned}	
\end{equation}

The scalar component ($\mathit{\Omega}^{\mathrm{S}}$) is essential for the trapping potential, but can be neglected when describing the spin-light interaction.

\subsubsection{Control laser and coherent control on the nuclear spin}
\label{sec:subcl}

The $\sigma^{+}$ circularly polarized control laser propagating along $\hat{x}$ has a polarization vector $\bm{u}=(0,1, i)/\sqrt{2}$. 
\begin{equation}
\label{circularcontrol}
\begin{aligned}
     H_{\mathrm{ac}} =\mathit{\Omega}_{x}^{\mathrm{V}} F_x+\mathit{\Omega}_{x,x}^{\mathrm{T}}F_x^2+\mathit{\Omega}_{y,y}^{\mathrm{T}}F_y^2+\mathit{\Omega}_{z,z}^{\mathrm{T}}F_z^2,  
\end{aligned}
\end{equation}
where $\mathit{\Omega}_{x}^{\mathrm{V}}=-\frac{\alpha_F^{\mathrm{V}}}{8F}\vert\mathcal{E}\vert^2$, $\mathit{\Omega}_{x,x}^{\mathrm{T}}=\frac{\alpha_F^{\mathrm{T}}}{4F(2F-1)}\vert\mathcal{E}\vert^2$, $\mathit{\Omega}_{y,y}^{\mathrm{T}}=-\frac{\alpha_F^{\mathrm{T}}}{8F(2F-1)}\vert\mathcal{E}\vert^2$, and $\mathit{\Omega}_{z,z}^{\mathrm{T}}=-\frac{\alpha_F^{\mathrm{T}}}{8F(2F-1)}\vert\mathcal{E}\vert^2$. Since there is an identity for the quadratic terms $F_x^2+F_y^2+F_z^2=F(F+1)I$, the uniform amplitudes in the three axes do not contribute to any rotational effect. We transform Eq.(\ref{circularcontrol}) to characterize the net spin rotational effect, $H_{\mathrm{c}}=\mathit{\Omega}_{x}^{(1)} F_x+\mathit{\Omega}_{x,x}^{(2)}F_x^2$, where $\mathit{\Omega}_{x}^{(1)}=-\frac{\alpha_F^{\mathrm{V}}}{8F}\vert\mathcal{E}\vert^2$, and $\mathit{\Omega}_{x,x}^{(2)}=\frac{3\alpha_F^{\mathrm{T}}}{8F(2F-1)}\vert\mathcal{E}\vert^2$.

The circularly polarized control laser propagates along the $\hat{x}$ direction. It has a beam diameter ($\sim$ 6.5 mm) much larger than the size of the atomic cloud ($\sim$ 100 \textmu m) to ensure homogeneous laser intensity for the spin-light interaction. Both the power and frequency of the control laser are actively stabilized. 
The calculated corresponding ground-state dynamical polarizabilities and generalized Rabi frequencies are plotted in Extended Data Fig.~\ref{fig5}.  When calculating the ground-state polarizabilities, only two upper levels, $\mathrm{6s6p\ ^1P_1}$ and $\mathrm{6s6p\ ^3P_1}$, are considered. The detunings of the control laser relative to the $\mathrm{^1S_0(F=5/2)\rightarrow ^3P_1(F'=3/2)}$ transition (Fig.~\ref{fig2}\textbf{b} and ~\ref{fig2}\textbf{c}) are -804 MHz and -962 MHz, respectively. The laser detunings relative to the hyperfine resonance are $10^3$ times the transition linewidth, and the probability of spontaneous emission during a Rabi period is less than 1\%. In Fig.~\ref{fig2}\textbf{b}, the extracted Rabi frequencies from the fits are $\mathit{\Omega}_{x}^{(1)}=-\mathit{\Omega}_{x,x}^{(2)}=2\pi\times0.5187(6)\ \mathrm{kHz}$. In Fig.~\ref{fig2}\textbf{c}, $\mathit{\Omega}_{x}^{(1)}=-\mathit{\Omega}_{x,x}^{(2)}/2=2\pi\times0.285(1)\ \mathrm{kHz}$.

\subsubsection{Optical lattice}

The optical lattice is linearly polarized along $\hat{z}$ and $\bm{u}=(0,0,1)$.
\begin{equation}
	H_{\mathrm{t}}= H_{\mathrm{ac}}=\widetilde{\mathit{\Omega}}_{z,z}^{(2)}F_z^2,
\end{equation}
where $\widetilde{\mathit{\Omega}}_{z,z}^{(2)}=-\frac{3\alpha_F^{\mathrm{T}}}{4F(2F-1)}\widetilde{\vert\mathcal{E}\vert}^2$. Here $\widetilde{\mathit{\Omega}}_{z,z}^{(2)}$ is derived in the same way as in section \ref{sec:subcl}. The field amplitude $\mathcal{E}$ is derived from the light intensity with the relation $\vert\mathcal{E}\vert^2=2I/(\varepsilon_0 c)$, and the symbol $\ \widetilde{ }\ $ denotes a physical quantity with probability distribution. The lattice beam has a wavelength of 1036 nm, a beam waist of 20 \textmu m and a trap depth of approximately 2.4 mK.

\subsection{Atomic-state preparation and state-selective measurement}

The atoms are spin-polarized by optical pumping in the $\hat{z}$ direction. The $\sigma^{+}$-polarized pump laser along the $\hat{z}$ direction is on resonance with the $\mathrm{^1S_0(F=5/2)\rightarrow ^1P_1(F'=5/2)}$ transition. After a 5-ms pump laser pulse, the atoms are optically pumped into the stretched state $\ket{F, F}$ ($\ket{5/2,+5/2}$).

The state-selective measurements are realized by absorption imaging along the $\hat{z}$ direction. The $\sigma^{+}$-polarized probe beam, co-propagating with the pump beam, is on resonance with the $\mathrm{^1S_0(F=5/2)\rightarrow ^3P_1(F'=7/2)}$ transition. We utilize the differential light shift imposed by the optical lattice to realize state-selective probing of atoms on the ground-state $\ket{5/2,+5/2}$. The differential polarizabilities between the ground state and probe state are
 \begin{equation}
\Delta\alpha(\lambda_{\mathrm{m}}, m)=\Delta\alpha^{(0)}(\lambda_{\mathrm{m}})+\Delta\alpha^{(2)}(\lambda_{\mathrm{m}})\frac{3 m^2-F(F+1)}{F(2F-1)}
\end{equation}
where $\Delta\alpha^{(0)}(\lambda_{\mathrm{m}})=\alpha^{(0)}({\mathrm{^3P_1}},\lambda_{\mathrm{m}})-\alpha^{(0)}({\mathrm{^1S_0}},\lambda_{\mathrm{m}})$, $\Delta\alpha^{(2)}=\alpha^{(2)}({\mathrm{^3P_1}},\lambda_{\mathrm{m}})$. The tensor polarizability of the ground state is neglected in this case. The probe laser couples the ground state ($\ket{m}_{F=5/2}$) with the probe state ($\ket{m+1}_{F'=7/2}$). At a `magic wavelength' of $\lambda_{\mathrm{m}}=1036$ nm, the differential polarizability of the $m=+7/2$ probe-state is zero, i.e., $\Delta\alpha(+7/2)=0$. Excitation to states except for the $m=+7/2$ probe state will be suppressed due to the nonzero differential light shift ($\Delta\nu\propto\Delta\alpha I$). At a trap depth of approximately 2.4 mK, the differential light shift is on the order of 2 MHz, much larger than the transition linewidth of 180 kHz.

\subsection{Pulse sequences}

Our experiment follows a generic sequence of initialization, coherent manipulation, and measurement of the spin state. The spin-state initialization is realized by optical pumping. The processes of coherent manipulation and spin-state measurement in different experimental demonstrations are given below.
\subsubsection{Spin-light interaction}\label{subsec5_1}
 \begin{enumerate}
\item A probe laser pulse is applied after the state initialization, and the number of probed atoms is $N$.
\item A control laser pulse is applied to drive the spin dynamics.
\item A probe laser pulse is applied for state-selective measurement of the atoms ($N_{+5/2}$) remaining in the $m=+5/2$ ground state, and the normalized population of the $m=+5/2$ ground state is $P_{+5/2}=N_{+5/2}/N$.
\end{enumerate}
\subsubsection{Ramsey spectroscopy}\label{subsec5_2}
 \begin{enumerate}
\item A $\pi/2$ pulse of the control laser is applied after the state initialization, and the spin state is driven into either the cat state or a CSS state.
\item After a Ramsey interrogation time of $\tau$, another $\pi/2$ pulse of the control laser is applied.
\item A probe laser pulse is applied for state-selective measurement of the atoms ($N_{+5/2}$) remaining in the $m=+5/2$ ground state.
\item A $\pi$ pulse of the control laser swaps the populations of the $m=+5/2$ and $m=-5/2$ ground state.
\item Then a second probe laser pulse is applied for state-selective measurement of the atoms ($N_{-5/2}$) transferred from $m=-5/2$ into the $m=+5/2$ ground state.
\item A pump laser pulse pumps all populations into the $m=+5/2$ ground state.
\item A probe laser pulse is applied to measure atoms ($N$) in the $m=+5/2$ ground state. 
The normalized population of the $m=\pm5/2$ ground state is $P_{\pm5/2}=N_{\pm5/2}/N$.
\end{enumerate}

\subsection{Magnetic field sensitivity}

We evaluate the magnetic field sensitivity with classical Fisher information\cite{Dietsche2019}. The single-shot magnetic field sensitivity set by the Cram{\'e}r-Rao bound is
\begin{equation}
\sigma_{B}^{1}=1/\sqrt{\mathcal{I}_{B}}
\end{equation}
where $\mathcal{I}_{B}$ is the classical Fisher information. $\mathcal{I}_B=(\frac{\partial \mathit{\Phi}}{\partial B})^2 \mathcal{I}_{\mathit{\Phi}}$ and
\begin{equation}
\mathcal{I}_{\mathit{\Phi}}=\sum_i \frac{\left(\partial P_i / \partial \mathit{\Phi}\right)^2}{P_i}
\end{equation}
where $P_{i}$ is the probability of outcome $i$, and $\mathit{\Phi}$ is the interferometric phase in Ramsey interferometry. After the Ramsey interrogation, only $m=\pm5/2$ ground-states are measured, then there are three different outcomes:
\begin{enumerate}
    \item   in the $m=+5/2$ state, $P_{+5/2}$,
    \item   in the $m=-5/2$ state, $P_{-5/2}$,
    \item  in intermediate states, $P_{\mathrm{in}}=1-P_{+5/2}-P_{-5/2}$.
\end{enumerate}

For the cat state, we fit $P_{\pm5/2}$ with cosine functions, 
\begin{equation}
P_{\pm 5/2}(\tau)=\frac{\overline{P_{\pm 5/2}}(\tau)}{2}[1+\mathcal{C}_{\pm 5/2}(\tau)\cos(\omega_{\pm 5/2}\tau+\varphi_{c,\pm 5/2})].
\end{equation}
The extracted coefficients show that $\overline{P_{+ 5/2}}(\tau)\approx\overline{P_{- 5/2}}(\tau)$, $\mathcal{C}_{+ 5/2}\approx\mathcal{C}_{- 5/2}$, $\omega_{+ 5/2}\approx\omega_{- 5/2}$, and $\varphi_{c,+ 5/2}\approx\varphi_{c,- 5/2}$. Then we take the average of these coefficients for $\overline{P}$ and $\mathcal{C}$ shown in Fig.~\ref{fig3}\textbf{c} and ~\ref{fig3}\textbf{d}. The interferometric phase $\mathit{\Phi}_{\mathrm{cat}}=\omega_{\pm 5/2} \tau=2F\gamma B \tau$. $\mathcal{I}_{\mathit{\Phi}}$ is maximal when $\mathit{\Phi}+\varphi=(n+1/2)\pi$. $\frac{\partial \mathit{\Phi}_{\mathrm{cat}}}{\partial B}=2F\gamma \tau$ and $(\mathcal{I}_{\mathit{\Phi}})_{\mathrm{max}}=\overline{P}\mathcal{C}^2$, then the single-shot magnetic field sensitivity of the cat state is
\begin{equation}
\sigma_{B,\mathrm{cat}}^{1}=\frac{1}{\mathcal{C}\sqrt{\overline{P}}} \frac{1}{2F} \frac{1}{\gamma \tau},
\end{equation}
where $2F$ is the quantum-enhanced factor. $\overline{P}=0.92(4)$ and $\mathcal{C}=0.82(5)$ after an interrogation time of 180 s and $\gamma(\mathrm{^{173}Yb})=-0.27196(1)\mu_{\mathrm{N}}/\hbar$\cite{Stone2005}. The corresponding magnetic field sensitivity is 0.11(1) nT.

For the CSS, we fit $P_{\pm5/2}$ with a cosine function consisting one fundamental frequency and two odd harmonics, 
\begin{equation}
\begin{aligned}
		P_{+5/2}&=\frac{\overline{P_{+5/2}}}{2}[1+\sum_{k}^{1,3,5}\mathcal{C}_{+5/2}^{(k)} \cos(k\mathit{\Phi}_{\pm 1/2}+\phi_{+5/2}^{(k)})],\\
		P_{-5/2}&=\frac{\overline{P_{-5/2}}}{2}[1-\sum_{k}^{1,3,5}\mathcal{C}_{-5/2}^{(k)} \cos(k\mathit{\Phi}_{\pm 1/2}+\phi_{-5/2}^{(k)})],\\
\end{aligned}
\end{equation}
where $\mathit{\Phi}_{\pm 1/2}=\omega_{\pm 1/2} \tau=\gamma B \tau$. The approximate results of $\mathcal{I}_{\mathit{\Phi}}$ is 
\begin{equation}
\begin{aligned}
		\mathcal{I}_{\mathit{\Phi}}&\approx \overline{P}\left[\sum_{k}^{1,3,5}k\mathcal{C}^{(k)} \sin(k\mathit{\Phi}+\phi^{(k)})\right]^2,\\
		&\approx\frac{\overline{P}}{(\gamma B)^2}\vert \frac{\partial \mathcal{P}_{z}(\tau)}{\partial \tau} \vert^2,
\end{aligned}
\end{equation}
where $\mathcal{P}_{z}=(P_{+5/2}-P_{-5/2})/(P_{+5/2}+P_{-5/2})$. The single-shot magnetic field sensitivity of the CSS is
\begin{equation}
\sigma_{B,\mathrm{css}}^{1}=\frac{1}{\vert\frac{\partial \mathcal{P}_{z}(\tau)}{\partial \tau} \vert_{\mathrm{max}}/(\gamma B)}\frac{1}{\sqrt{\overline{P}}}\frac{1}{\gamma \tau}.
\end{equation}

Due to dephasing between several decoherence-free subspaces, the CSS evolves into $\rho_{\mathrm{mix}}$. The weighting factors $w_{\pm m}$ in $\rho_{\mathrm{mix}}$ are
\begin{equation}
	w_{\pm m}=\frac{1}{2^{2F-1}}\binom{2F}{F+\vert m\vert}.
\end{equation}
The second $\pi/2$ pulse of the control laser recombines the superposition. It converts the phase accrued during $\tau$ to a population imbalance among ground states, and $\rho_{\mathrm{mix}}^{*}=\mathrm{e}^{-\mathrm{i}F_x\pi/2}\rho_{\mathrm{mix}}\mathrm{e}^{\mathrm{i}F_x\pi/2}$ afterward. Populations are lost in intermediate states and $P_{F}=\braket{  F,F\vert \rho_{\mathrm{mix}}^{*} \vert F,F}= \braket{ \pi/2,\pi/2\vert \rho_{\mathrm{mix}}\vert \pi/2,\pi/2}$. For $F=5/2$, the theoretical prediction gives $P_{+5/2}+P_{-5/2}=0.4922$. After an interrogation time of 180 s, $\overline{P}=0.51(2)$, consistent with the theoretical prediction.

\subsection{Units and errors}\label{subsec8}

In this article, all errors of repeated measurements and fitted results denote a 1-s.d. confidence interval. The trap lifetime is the 1/e decay time. 
\backmatter 



\bmhead{Acknowledgments}
We thank D. Sheng and M. Krstaj{\'i} for fruitful discussions, and W.-K. Hu for contribution to the apparatus in the early stage. 
\bmhead{Funding}
This work has been supported by the National Natural Science Foundation of China (NSFC) through Grants No. 12174371 and the Innovation Program for Quantum Science and Technology through Grant No. 2021ZD0303101. C.-L.Z. was supported by NSFC through Grant No. 11922411, and the Innovation Program for Quantum Science and Technology through Grant No. 2021ZD0300203.
\bmhead{Competing interests}
The authors declare no competing interests.

\bmhead{Author contributions}
All authors contributed to carrying out the experiments, interpreting the results, and writing the manuscript.	

\bibliography{sn-bibliography2}

\clearpage
\begin{appendices}
\section*{Extended Data Figures}

\renewcommand\figurename{Extended Data Fig.}

\begin{figure}[htb]
\centering
\includegraphics[width=1.0\textwidth]{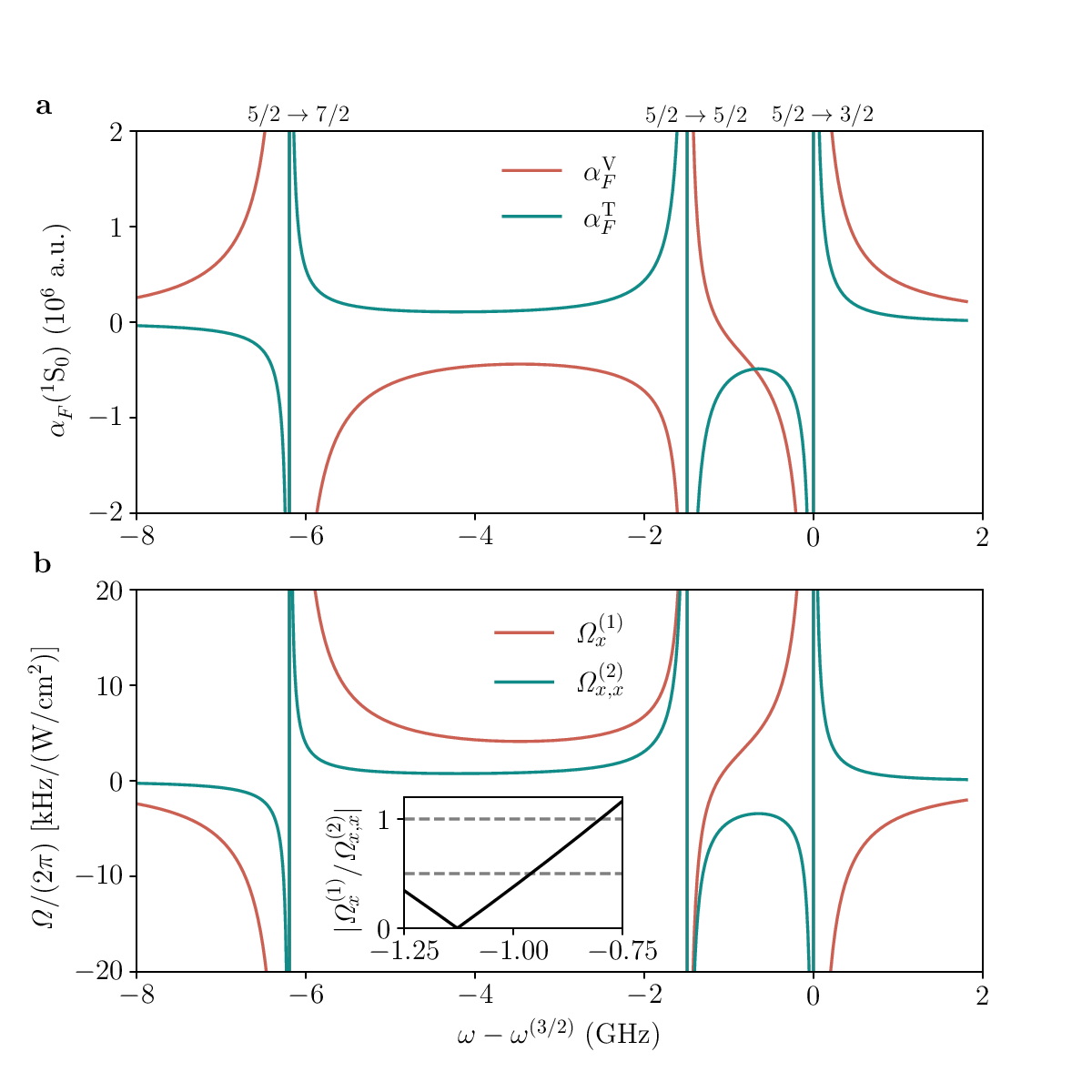}
\caption{\textbf{a}, The dynamical vector and tensor polarizabilities $\alpha_F^{\{\mathrm{V,T}\}}$ of $\mathrm{^{1}S_{0}}$ are plotted as a function of the laser frequency, whose value is relative to the $\mathrm{6s^2\ ^1S_0}\rightarrow \mathrm{6s6p\ ^3P_1},\ F=5/2 \rightarrow F=3/2$ hyperfine transition. For polarizabilities 1 a.u. = $\mathrm{1.648\ 773\times 10^{-41}\ C\ m^2\ V^{-1}}$. \textbf{b}, The generalized Rabi frequencies $\mathit{\Omega}_{x}^{(1)}$ and $\mathit{\Omega}_{x,x}^{(2)}$ are plotted as a function of the laser frequency, where $\mathit{\Omega}_{x}^{(1)}=-\frac{\alpha_F^{\mathrm{V}}}{8F}\vert\mathcal{E}\vert^2$ and $\mathit{\Omega}_{x,x}^{(2)}=\frac{3\alpha_F^{\mathrm{T}}}{8F(2F-1)}\vert\mathcal{E}\vert^2$. The Rabi frequencies are normalized assuming a light intensity $I$ of $\mathrm{=1\ W/cm^2}$, and $\vert\mathcal{E}\vert^2=\frac{2I}{\epsilon_0 c}$. The inset shows the frequency ratio $\vert\mathit{\Omega}_{x}^{(1)}/\mathit{\Omega}_{x,x}^{(2)}\vert$.}\label{fig5}
\end{figure}
\end{appendices}

\end{document}